\documentclass[aps,pre,reprint,twocolumn,superscriptaddress,showkeys]{revtex4-2}

\usepackage{amsmath}
\usepackage{amssymb}
\usepackage{graphicx}
\usepackage{bm}
\usepackage{hyperref}
\usepackage{float}

\begin{document}

\title{Analysis of Josephson junctions switching time distributions for the detection of single microwave photons}

\author{A. S. Piedjou Komnang}
\affiliation{Dipartimento di Fisica ``E.R. Caianiello'', Universit\`a di Salerno, Via Giovanni Paolo II, 132, Fisciano I-84084 SA, Italy}

\author{C. Guarcello}
\email[Corresponding author: ]{cguarcello@unisa.it}
\affiliation{Dipartimento di Fisica ``E.R. Caianiello'', Universit\`a di Salerno, Via Giovanni Paolo II, 132, Fisciano I-84084 SA, Italy}
\affiliation{INFN, Gruppo Collegato Salerno, Fisciano I-84084 SA, Italy}

\author{C. Barone}
\affiliation{Dipartimento di Fisica ``E.R. Caianiello'', Universit\`a di Salerno, Via Giovanni Paolo II, 132, Fisciano I-84084 SA, Italy}
\affiliation{INFN, Gruppo Collegato Salerno, Fisciano I-84084 SA, Italy}

\author{C. Gatti}
\affiliation{INFN, Laboratori Nazionali di Frascati, Frascati (Roma) Italy and Dept. of Mathematics and Physics, University of Roma Tre, Roma I-00100, Italy}

\author{S. Pagano}
\affiliation{Dipartimento di Fisica ``E.R. Caianiello'', Universit\`a di Salerno, Via Giovanni Paolo II, 132, Fisciano I-84084 SA, Italy}
\affiliation{INFN, Gruppo Collegato Salerno, Fisciano I-84084 SA, Italy}

\author{V. Pierro}
\affiliation{Dept. of Sciences and Technologies, Univ. of Sannio, Benevento I-82100, Italy}
\affiliation{INFN, Gruppo Collegato Salerno, Fisciano I-84084 SA, Italy}

\author{A. Rettaroli}
\affiliation{INFN, Laboratori Nazionali di Frascati, Frascati (Roma) Italy and Dept. of Mathematics and Physics, University of Roma Tre, Roma I-00100, Italy}

\author{G. Filatrella}
\affiliation{Dept. of Engineering, University of Sannio, Benevento I-82100, Italy}
\affiliation{INFN, Gruppo Collegato Salerno, Fisciano I-84084 SA, Italy}

\date{Received 30 September 2020; Revised 13 November 2020; Accepted 18 November 2020; Available online 28 November 2020}

\begin{abstract}
We investigate an optimal scheme for the detection of single microwave photons by a Josephson junction through the analysis of its switching times distribution. The proposed analysis is of support for the decision about the existence of the photon field, which is important in the case of rare events. We assume that the cavity and the transmission line are ideal (each photon absorbed to the cavity gives a current pulse as the output of the transmission line) and the photon source is periodic. The employed methodology consists in comparing the switching probabilities of a Josephson junction exposed to a train of current pulses, simulating a weak photon field, with that of the same device in absence of pulses. In both cases, thermal noise can induce thermal activated switchings. The investigation of the unbalance in the number of switching events in the two cases, gives an estimate of the efficiency of the detection. Furthermore, in the assumption of escapes described by Kramers model, it is possible to provide a relationship between the properties of the photons field, the quantum efficiency of the detection process, and the Josephson junctions switching features at finite temperatures.
\end{abstract}


\maketitle

Josephson junctions (JJs) can be interesting as radiation detectors, as they can reach the quantum sensitivity limit \cite{guarcello2017anomalous,guarcello2019josephson}. Moreover, as they operate at cryogenic temperatures, intrinsic noise can be reduced as much as cryogenics allows. On this basis, JJs are very promising for the detection of very weak electromagnetic signals in the microwave spectral region, possibly close to the single photon limit \cite{anghel2020cold,brange2018nanoscale,chen2011microwave,guarcello2019nonlinear,guarcello2019calorimeter,karimi2020quantum,leoni2006fabrication,oelsner2017detection,oelsner2013underdamped,poudel2012quantum,revin2020microwave,walsh2017graphene}. This can be of relevance in basic physics experiments, such as axion detection \cite{alesini2020status,alesini2020development,asztalos2010squid,beck2011testing,beck2013possible,beck2015axion,beck2016cosmological,dixit2018detecting,kahn2016broadband,kuzmin2018single,matlashov2018squid,popov2016resonance,yan2020nonlinear}. The detection of photons calls for a realistic model describing the coupling between the electromagnetic resonator (the cavity absorbing the photon) and a JJ. Such a setting is complicated by the fact that both systems, i.e., the cavity and the junction, should be treated as quantum damped oscillators \cite{anghel2020electromagnetic,caldeira1981influence,peropadre2011approaching,poudel2012quantum,schondorf2018optimizing}. In our approach, we make the simplifying assumptions that both the cavity and the transmission line are ideal, i.e., each absorbed photon produces a current pulse that is injected into the junction as the output of the transmission line (thus implicitly neglecting any interaction between the cavity and the JJ, e.g., of the type described in Ref.~\cite{shukrinov2016modeling}), and the photon source is periodic. However, the practical realization of Josephson based single microwave photon detectors may encounter another difficulty connected to the specific nature of the superconducting phase, as the response of the JJ to the photon perturbation has to be revealed through the passage of the JJ from the static to the dynamic state; this passage is signaled by the appearance of a voltage. Further complications are introduced by the nonlinearity of the phase confining potential, and especially by the disturbances due to intrinsic noise and competing quantum tunneling effects. A principal question is therefore how to discern the photon arrival, given the limited information available (the appearance of a voltage) and the presence of noise. The response of a JJ to a train of current pulses is analyzed in the frame of signal detection theory \cite{addesso2012characterization,filatrella2010detection}. The ultimate aim of this line of thinking is the detection of single photons in the presence of noise. The optimization of the detection probability (and the minimization of the false signal probability) also provide a guide for the selection of the JJ parameters that best suite specific detection applications. More specifically, in Section~1 the used JJ model is described, the methods to analyze the switching events are developed in Section~2, and in Section~3 the resulting performances of the sketched device are collected. Finally, Section~4 summarizes the main findings and outlook.

\section{Model of the Josephson junction detector}
The electrodynamics of a small sized, superconductor--insulator--superconductor (SIS) type JJ can be described in terms of the well known Resistively and Capacitively Shunted Junction (RCSJ) model \cite{barone1982physics,spagnolo2017nonlinear,spagnolo2015noise}, shown in Fig.~1, an equivalent electrical model whose circuit elements are related to specific junction physical characteristics. With reference to Fig.~1, the capacitor $C_J$ represents the capacitance between the junction electrodes, the resistor $R_J$ the conduction path due to the tunneling of normal electrons (quasiparticles), and the element JJ the conduction path of superconductive electrons (Cooper pairs). The current through the JJ element and the corresponding voltage drop are related to the gauge invariant phase difference, $\phi$, between the macroscopic wavefunctions of the two superconductors by the well known Josephson equations:
\begin{equation}
I_J = I_0 \sin \phi,
\end{equation}
\begin{equation}
V = \frac{\hbar}{2e} \frac{d\phi}{dt}.
\end{equation}

\begin{figure}[htbp]
\centering
\includegraphics[width=\columnwidth]{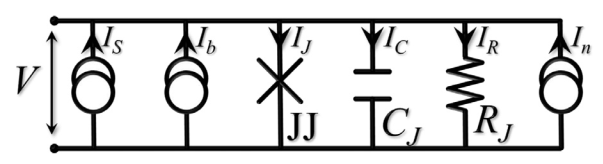}
\caption{Electrical model of a detector based on a Josephson junction.}
\label{fig:fig1}
\end{figure}

Here, $I_0$ is the maximum Cooper pair current that can flow in the JJ element, and $\hbar$ and $e$ are the reduced Planck constant and elementary charge, respectively. When a dc current below $I_0$ flows in the JJ element, the Josephson equations predict a possible solution with a stationary phase and a zero average voltage. If the dc bias current overcomes $I_0$, the same equations predict a switching to a finite voltage state. It is worth noting that, in a typical SIS JJ, the resistor $R_J$ depends quite strongly both on voltage and temperature. However, as the overall effect of the resistor is to introduce dissipation in the system, its nonlinearity is often not considered if the damping is moderate or weak. Another important effect of the quasiparticle current, described by $R_J$, is the introduction of a noise current source, through the fluctuation dissipation theorem, whose spectral power density is assumed to be frequency independent (Johnson noise) \cite{barone2018comparison,kogan1996electronic}. Such noise current source is indicated with $I_n$ in Fig.~1.

In order to use a JJ as a detector, it is convenient to bias it with a dc current just below $I_0$, indicated with $I_b$ in Fig.~1, so that the occurrence of an external disturbance (i.e., the signal to be detected) can switch the junction from the zero to the finite voltage state. Of course the current noise can also induce a switching, so appropriate measures have to be taken to distinguish proper detection from false alarms (dark counts, in the detectors jargon). We observe that other different detection strategies can be adopted as well, e.g, biasing the junction with an ac electric current \cite{addesso2012characterization,filatrella2010detection,guarcello2013role,guarcello2015phase,pountougnigni2020detection,yablokov2020suppression}.

To model a weak microwave field (weak because carrying few photons) coupled to the junction, we consider a deterministic current source, indicated with $I_s$ in Fig.~1, consisting of a train of well separated periodic current pulses, each representing the current injected by the absorption of a single microwave photon, shown in Fig.~2. We make the further assumption that each photon-induced current pulse is rectangular.

\begin{figure}[htbp]
\centering
\includegraphics[width=\columnwidth]{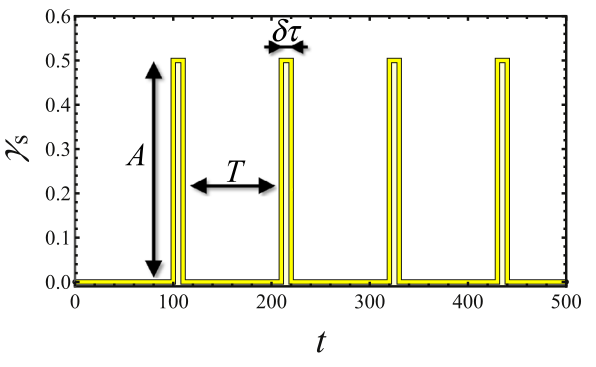}
\caption{Train of pulses that mimics the effect of photons absorption.}
\label{fig:fig2}
\end{figure}

Denoting with $I_J$ the current through the JJ element, $I_R$ the current through the resistor, and $I_C$ the current through the junction capacitance, it is possible to write the current balance equation:
\begin{equation}
I_C + I_R + I_J = I_b + I_s(t) + I_n(t).
\end{equation}
By using the constitutive relations of the resistor and the capacitor, and the Josephson relations, it is straightforward to write the following second order differential equation:
\begin{equation}
C_J \frac{\hbar}{2e}\frac{d^2\phi}{dt^2} + \frac{1}{R_J}\frac{\hbar}{2e}\frac{d\phi}{dt} + I_0 \sin\phi = I_b + I_s(t) + I_n(t),
\end{equation}
which, due to the presence of the stochastic noise term $I_n$, is a Langevin equation \cite{benjacob1984thermal}. By defining a normalized time $\tau = \omega_J t$, where $\omega_J = \sqrt{2eI_0/C_J\hbar}$ is the Josephson plasma frequency \cite{barone1982physics}, Eq.~(4) can be rewritten as:
\begin{equation}
\frac{d^2\phi}{d\tau^2} + \alpha_J \frac{d\phi}{d\tau} + \sin\phi = \gamma_b + \gamma_s(\tau) + \gamma_n(\tau),
\end{equation}
where the parameters are defined as
\begin{equation}
\alpha_J = \frac{1}{R_J C_J \omega_J}, \quad \gamma_b = \frac{I_b}{I_0}, \quad \gamma_s = \frac{I_s}{I_0}.
\end{equation}
The statistical properties of the noise term $\gamma_n = I_n / I_0$ are:
\begin{equation}
\langle \gamma_n(\tau) \rangle = 0, \quad \langle \gamma_n(\tau), \gamma_n(\tau') \rangle = 4D\delta(\tau - \tau'),
\end{equation}
where $D = k_B T \omega_J / (R_J I_0^2)$ is the normalized noise intensity ($k_B$ is the Boltzmann constant and $T$ the absolute temperature), $\delta(\cdot)$ is the Dirac delta function, and the parenthesis $\langle \cdot \rangle$ represent ensamble averages.

As stated before, a train of current pulses is used to mimic the arrival of photons, as:
\begin{equation}
\gamma_s(\tau) = \sum_{n=-\infty}^{+\infty} [+A\theta(\tau - nT) - A\theta(\tau - n(T+\delta\tau))],
\end{equation}
where $\theta$ is the Heaviside step function, $A$ is the amplitude of the signal, $\delta\tau$ is the pulse width, and $T$ is the distance between two consecutive pulses. An example of a pulse train is shown in Fig.~2. The parameters used are: $A = 0.5$, $\delta\tau = 10$, and $T = 100$.

In absence of dissipation and noise, the dynamics of Eq.~(5) can be described in terms of a potential energy, having the shape of a washboard, with an infinite sequence of local minima where the phase can be trapped, giving rise to a zero average voltage state. Given enough energy, the phase can leave the local minima and run down the slope of the potential, generating a finite voltage state, as evident from Eq.~(2). The tilted washboard potential has the following functional form \cite{barone1982physics,benjacob1984thermal}:
\begin{equation}
U(\phi) = 1 - \cos(\phi) - \gamma\phi,
\end{equation}
with a slope given by the current $\gamma$ flowing through the junction, see Fig.~3. In the same figure is represented, as a small ball, the value of the JJ phase trapped in a potential well. If the phase overcomes the critical value $\phi^*$, a switch to the running state occurs. For $\gamma < 1$ the potential has metastable wells with a barrier height \cite{benjacob1984thermal}:
\begin{equation}
\Delta U(\gamma) = 2 \left[ \sqrt{1 - \gamma^2} - \gamma \cos^{-1}(\gamma) \right].
\end{equation}

\begin{figure}[htbp]
\centering
\includegraphics[width=\columnwidth]{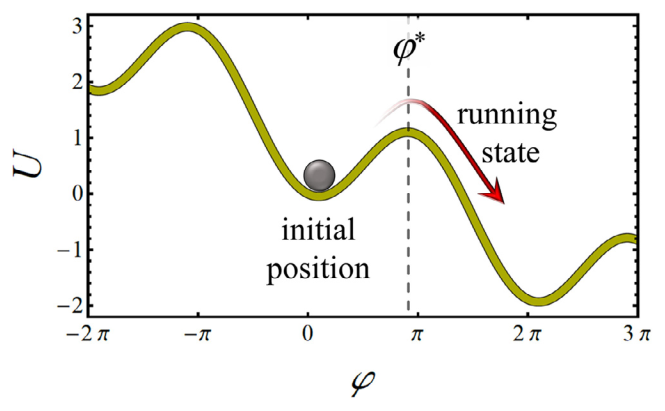}
\caption{Sketch of the tilted washboard potential. The initial state, or position, of the system, when the particle resides in a metastable well of the potential, is also depicted.}
\label{fig:fig3}
\end{figure}

In presence of noise the phase solution in the potential minima is metastable, and the resulting escape rate $r(\gamma)$ is given by the Kramers approximation for moderate damping \cite{kramers1940brownian}. By defining an escape attempt rate as:
\begin{equation}
\omega_0(\gamma) = (1 - \gamma^2)^{1/4}
\end{equation}
and the (imaginary) frequency of the oscillations, at the potential maximum, as $\omega_b$, it is possible to write the escape rate from the confining barrier (10) as:
\begin{equation}
r_0(\alpha_J, \gamma, D) = \frac{\omega_0}{\omega_b} \left( \sqrt{\frac{\alpha_J^2}{4} + \omega_b^2} - \frac{\alpha_J}{2} \right) \frac{e^{-\frac{\Delta U(\gamma)}{D}}}{2\pi}.
\end{equation}
The prefactor depends on the dissipation and the approximation employed \cite{risken1989fokker}; it can be simplified assuming that the two frequencies, at the top and bottom of the barrier, are identical, $\omega_0 = \omega_b$, so that the escape rate becomes:
\begin{equation}
r_0(\alpha_J, \gamma, D) = \left( \sqrt{\frac{\alpha_J^2}{4} + \omega_0^2} - \frac{\alpha_J}{2} \right) \frac{e^{-\frac{\Delta U(\gamma)}{D}}}{2\pi}.
\end{equation}
In Eq.~(12) the major contribution arises from the bias dependent barrier height \cite{barone1982physics}. The dependence on the dissipation is weak, and to a good approximation the rate can be simplified into:
\begin{equation}
r_0(\gamma, D) = \frac{\omega_0}{2\pi} e^{-\frac{\Delta U(\gamma)}{D}}.
\end{equation}
When the junction is dc biased, so that the phase resides in a local minimum (zero voltage state), the presence of thermal noise reduces the lifetime of such state towards the switching to the finite voltage state. In absence of external signals such escapes occur randomly, with the usual Kramers rate (12) \cite{benjacob1984thermal}.

In summary, JJs can be seen as threshold detectors that provide a promptly measurable signal (the voltage state) when subject to a disturbance whose energy is comparable with a characteristic threshold, i.e., the energy barrier height given in Eq.~(10). The unavoidable presence of intrinsic noise determines random switchings with a certain statistical distribution. The absorption of a photon adds to the noise and modifies the switching distribution. The analysis of the change in such distribution, as well as the methods to detect the presence of induced photons, is the subject of the present work.

\section{Detection}
In this Section are analyzed some general considerations about the possibility to retrieve information on the detection of weak signals through the analysis of the distribution of the JJ switching times.

\subsection{General features}
A simple calculation for the detection of a train of photons of frequency $\nu$ (and hence of energy $h\nu$) at the rate $r_A$ is the following. Assuming that the JJ is biased in such a way that the energy barrier height given in Eq.~(10) matches the photon energy $h\nu$, and that the small amplitude oscillation frequency of the phase particle in the bottom of the potential well, which is equal to the attempt rate (11), resonates with the photon frequency, the Kramers escape (14) (in the absence of the photons) reads:
\begin{equation}
r_0 = \nu \exp \left( - \frac{h\nu}{k_B T} \right).
\end{equation}
If one imposes that the thermal activation rate, given by Eq.~(15), matches the rate of photon arrivals, that is
\begin{equation}
r_A \approx r_0,
\end{equation}
straightforward algebraic manipulation of Eq.~(16) gives:
\begin{equation}
T \approx \frac{h\nu}{k_B} \left/ \log\left(\frac{\nu}{r_A}\right) \right..
\end{equation}
In case of detection of rare photons, e.g., with an arrival rate $r_A \approx 0.001\text{ Hz}$ and having a frequency in the microwave region (say $\nu \approx 14\text{ GHz}$), Eq.~(17) provides an operating temperature of $T^* \approx 22\text{ mK}$. At this level of approximation, that is exclusively considering thermal effects, detection of single microwave photon depends only on its frequency and arrival rate, and ultra-low temperatures would be necessary. A further source of intrinsic, temperature-independent switchings is introduced by macroscopic quantum tunneling (MQT) processes \cite{clarke1988quantum}. The rate due to quantum tunneling, $r_{\text{MQT}}$, could be embodied as a contribution, possibly the most relevant, to the intrinsic rate, $r_{\text{int}} = r_0(T) + r_{\text{MQT}}$, that should play the role of $r_0$ in Eq.~(16). However, for the sake of simplicity, quantum effects \cite{anghel2020electromagnetic,falci2013design,schondorf2018optimizing} are kept outside the purposes of this work.

It is useful to critically examine the estimate in Eq.~(17). The matching condition implies that one half of the detected JJ switchings is due to the absorption of photons and the other half are thermally induced. Therefore, at temperature below $T^*$ most of the escapes are due to photons, and simple counting is adequate to identify the presence of photons. However, reaching and maintaining a temperature as low as $T^*$ is very demanding from the technological point of view. It is therefore important to carefully analyze the data to decide about the presence of the photons also in region above the temperature $T^*$, when the number of extra photons is comparable to the number of thermal activated switches, as well as to quantify the reliability of the detection. An analysis based on signal detection theory is the subject of next Section.

\subsection{Statistical detection analysis}
In the context of signal detection \cite{helstrom1994elements,levine1973fondements}, one can be interested in determining the arrival time of the photon, or just in knowing whether or not a photon has been detected. In the latter case, only a limited information is interesting and, therefore, it is possible to optimize the statistical decision, at the cost of sacrificing the determination of the arrival time.

To start with, let us assume that a JJ has been repeatedly prepared (we assume that the number of repetitions is $N$) in the initial position as shown in Fig.~3. Let us also assume that each measurement session lasts a time $P$, during which there occur passages from the superconducting state to the running state at finite voltage. In this manner one can record, for a certain number $N$ of escape events, the switching times, that is the time the JJ takes to pass from the zero to the finite voltage state, $\bm{\tau} = \{\tau_i\}_{i=1}^N$.\footnote{In this description, we neglect the reset time -- that is the time to reallocate the JJ to the initial position from the running state. As electronic is usually very fast, this time is small compare to the waiting time between two consecutive switching events.} The switching times can be visualized in the form of a histogram. Qualitatively, it is reasonable to ask whether the retrieved histogram is compatible with the presence of thermal noise alone (hypothesis $H_0$), or with the contemporary absorption of microwave photons, modeled by a train of current pulses, Eq.~(8) (hypothesis $H_1$). The problem can be represented in Fig.~4, considering the two hypothesis of the binary test as:
\begin{align*}
H_0 &\equiv \{\text{Null hypothesis}\} \\
    &\equiv \{\text{Only thermal noise induces escapes}\}
\end{align*}
and
\begin{align*}
H_1 &\equiv \{\text{Alternative hypothesis}\} \\
    &\equiv \{\text{Thermal noise and the arrival of current pulses} \\
    &\qquad \text{at a rate } r_A \text{ induce escapes}\}.
\end{align*}

\begin{figure}[htbp]
\centering
\includegraphics[width=\columnwidth]{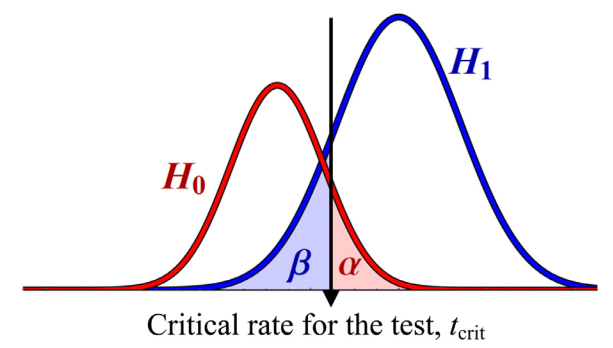}
\caption{Sketch of the decision scheme.}
\label{fig:fig4}
\end{figure}

The distribution under the null hypothesis $H_0$, the red curve in Fig.~4, is the distribution of the escapes in the case of purely thermally activated switches. The blue curve describes the distribution under the alternative hypothesis $H_1$ of the combined effect of thermally activated plus photon-induced escapes. The decision is taken with a somewhat arbitrary threshold $t_{\text{crit}}$. The quantity $\alpha$ denotes the probability of false detection: the detection of photons while the switchings are produced by thermal fluctuations. Analogously, $\beta$ is the probability to miss the signal: the photons are not detected and are confused with thermally activated switchings. The lower the quantities $\alpha$ and $\beta$, the better the test. The proper choice of the $t_{\text{crit}}$ value should provide both a low significance, e.g., $\alpha \le 1\%$ to exclude that the pure thermal noise explains the results, and a sufficiently high test power, e.g., $1 - \beta \ge 99\%$ to guarantee that the presence of a signal has not gone amiss. Thus the values of the quantities $\alpha$ and $\beta$ depend upon the application. In case of rare particles experiments, where indications about the possible presence of a particle is important, also relatively high $\alpha$ and $\beta$ values could be considered. Conversely, in telecommunication, where the acceptable error rate is very low, the values of $\alpha$ and $\beta$ must be much lower. To provide a flexible tool to determine the detection, a convenient way to quantify the discrimination power is the Kumar-Caroll (KC) index $d_{\text{KC}}$ \cite{kumar1984performance}, that has been already employed for the detection of sinusoidal signals through the analysis of the escape times of JJ \cite{addesso2012characterization,filatrella2010detection}:
\begin{equation}
d_{\text{KC}} = \frac{|\langle \tau_{\text{sw}} \rangle_1 - \langle \tau_{\text{sw}} \rangle_0|}{\sqrt{\frac{1}{2}\left[\sigma^2(\tau_{\text{sw}})_1 + \sigma^2(\tau_{\text{sw}})_0\right]}},
\end{equation}
where
\begin{equation}
\langle \tau_{\text{sw}} \rangle_1 = \left. \frac{1}{N}\sum_{i=1}^N \tau_i \right|_1
\end{equation}
is the estimated average switching time in the presence of the signal, and
\begin{align}
\sigma^2(\tau_{\text{sw}})_1 &= \text{Var}[\langle \tau_{\text{sw}} \rangle_1] = \left. \text{Var}\left[\frac{\tau_1 + \tau_2 + \dots + \tau_N}{N}\right] \right|_1 \nonumber \\
&= \frac{1}{N(N-1)}\sum_{i=1}^N (\tau_i - \langle \tau_{\text{sw}} \rangle_1)^2
\end{align}
is the estimate of the variance of the average switching time. The same quantities with the subscript 0 identify the values in the absence of the signal (here, it is important to stress the fact that all averages are taken over noise realizations). The KC index defined by Eq.~(18) is a qualitative indication, depending on the specific statistical distribution, and corresponds to the Signal-to-Noise-Ratio (SNR) only in the case of Gaussian distributions. However, one can use the quantity $d_{\text{KC}}$ as a first estimate of the capability to discriminate the two conditions (with or without photon-induced switches). Indeed, by comparing the average (19) (more accurately, the sample mean) with the corresponding average in the absence of external signal, the index is clearly related to the $t$-test statistics. Although this is a good indicator, it is clear, however, that a full test is needed to compare the entire information content of the escape times distribution and, therefore, usually can outperform the $t$-test \cite{addesso2012characterization}.

\subsection{Dichotomous classification of the switching events}
As described in the previous Section, a typical experiment consists of a run in which one waits for an overall observation time $P$ during which the switching events are recorded. In the above analysis we have assumed that all switching events occur short of the observation time $P$. Under this hypothesis, the switching distribution is continuous and the KC index (18) measures the goodness of the detection. If instead the hypothesis of switches within $P$ is released, there will be a portion of the measurement runs that will not exhibit switching events. As a consequence, the collection of switching times might be relatively poor and anyway deformed, for a portion of the switching times, those longer than $P$, are not registered. In these circumstances, it is conceivable to detect the signal -- or to distinguish whether the distribution of the switches is drawn from the purely noise distribution or from the distribution of the noise added to a train of pulses -- grouping the measurement runs in two classes: those in which at least an escape is registered, and those in which it is not. This dichotomous classification of the experimental runs can be dealt with a counting statistics. To do so, one defines the probabilities, $p_0$ and $p_1$, of a switch to occur prior to the wall-time $P$ under the two hypothesis $H_0$ and $H_1$, respectively:
\begin{align}
p_0 &= P(\tau_i \le P | H_0) \\
p_1 &= P(\tau_i \le P | H_1).
\end{align}
The above probabilities are to be estimated from the experiments. We denote with a $\sim$ the estimates for a finite $N$:
\begin{align}
\tilde{p}_0 &\approx \frac{N_0}{N}, \quad N_0 \equiv \# \tau_i \le P | H_0 \\
\tilde{p}_1 &\approx \frac{N_1}{N}, \quad N_1 \equiv \# \tau_i \le P | H_1.
\end{align}
One can recognize that the estimates (22) are Gaussianly distributed with variance $\approx p(1 - p)$. It is therefore possible to estimate the SNR through the following $d_{\text{KC}}$ index:
\begin{equation}
d_{\text{KC}} = \sqrt{N} \frac{|\tilde{p}_1 - \tilde{p}_0|}{\sqrt{\frac{1}{2}\left[\tilde{p}_1(1 - \tilde{p}_1) + \tilde{p}_0(1 - \tilde{p}_0)\right]}}.
\end{equation}
If a switching occurs in any run, the probabilities $\tilde{p}_1$ and $\tilde{p}_0$ read $\tilde{p}_1 \approx \tilde{p}_0 \approx 1$, and the SNR with this method vanishes. Therefore, the detection method described in this Section is to be pursued if some, but not all, of the experimental runs show switching events \cite{filatrella2020analysis}.

\section{Detection optimization}
In this Section we optimize the case in which a sizable number of switches occurs during each experimental run, as described in Section~2.2.

The lay-out of the problem of previous Section allows some analytical estimates of the performances. It is useful to estimate the $d_{\text{KC}}$ index dependence upon the length of the observation period. The number of purely thermal escapes in the observation time $P$ is:
\begin{equation}
N_0 \equiv \{\#\text{ thermal activated switches}\} = r_0 P,
\end{equation}
while the number of switches in the presence of the photon field is:
\begin{align}
N_1 &\equiv \{(\#\text{thermal activated} + \#\text{photon-induced})\text{switches}\} \nonumber \\
&= r_1 P.
\end{align}
For an estimate of the detection features, let us assume that the escape in the observation time $P$ is a Poisson process. If this is the case, the number of counts variance coincides with the average number of counts. Therefore, it is possible to estimate the quantities in (18) through the Kramers rate for moderate damping (12) and an additional term for the influence of the photons on the escapes. An estimate of $N_1$ can be achieved adding the photon rate to the Kramers' rate (12) for the purely thermal case:
\begin{equation}
r_1 \approx r_0(\gamma, D) + r_A.
\end{equation}
The approximation (26) enables to estimate the number of escapes (24) and (25) and the variances of the sample means:
\begin{align}
\sigma^2(\tau_{\text{sw}})_0 &\approx N_0 \\
\sigma^2(\tau_{\text{sw}})_1 &\approx N_1.
\end{align}
The above estimates, after straightforward algebraic manipulations, give for Eq.~(18):
\begin{equation}
d_{\text{KC}} = \frac{r_A P}{\sqrt{\frac{1}{2}[(r_A + r_0)P + r_0 P]}} = \sqrt{2P}\frac{r_A}{\sqrt{r_A + 2r_0}}.
\end{equation}
Assuming that a reasonable threshold for reliable detection is $d_{\text{KC}} \ge 1$, the previous equation becomes:
\begin{equation}
P r_A^2 - \frac{1}{2}r_A - r_0 \ge 0,
\end{equation}
that implicitly defines the connection between the arrival rate of the photons $r_A$, the thermal escape rate $r_0$, and the observation time $P$. An immediate result from (28) is the following: once the photon arrival and the thermally induced switching rates are given, an increase of the observation time $P$ provides a more reliable detection, as the SNR estimate grows with the square root of $P$.

Some elaborations of the estimate (29) are relevant. By defining the ratio between the spontaneous thermal escapes and the photon arrival rate as:
\begin{equation}
x = \frac{r_0}{r_A}
\end{equation}
it is possible to observe that:
\begin{equation}
d_{\text{KC}} = \sqrt{\frac{2Pr_A}{2x + 1}}.
\end{equation}
This equation confirms the intuitive scaling between the observation time $P$, the photon arrival rate, and the ratio between the thermal and photon rates. In particular, one can insert the matching condition (16) $P = 1/r_A$ in Eq.~(31) to obtain a relation between the observation time, the temperature, and the bias point through Eqs.~(11)--(14).

Another interesting consideration is that the number of photons that actually cause a switching of the JJ, $r_d P$, does not necessarily coincide with the number of incoming photons $r_A P$. One can define the quantum efficiency $\eta$ that counts the fraction of photon actually detected by the JJ:
\begin{equation}
r_d = \eta r_A.
\end{equation}
At this point, it is quite natural to write the efficiency as a function of the parameters, that are again the bias current and the noise intensity. If one writes explicitly such dependence $\eta = \eta(\gamma, D)$, Eq.~(29) for the observation time $P_\eta$ becomes:
\begin{equation}
P_\eta [\eta(\gamma, D) r_A]^2 - \frac{1}{2}\eta(\gamma, D)r_A - r_0(\gamma, D) = 0,
\end{equation}
so that
\begin{equation}
P_\eta = \frac{\eta(\gamma, D)r_A + 2r_0(\gamma, D)}{2[\eta(\gamma, D)r_A]^2}.
\end{equation}
In this estimate, the observation time $P_\eta$ is a function of $\eta$ and $r_0$. The quantum efficiency can be numerically, or experimentally, evaluated also using a source of photons with a relatively high rate $r_A$, as long as the events are sufficiently separated so that the system relaxes to equilibrium between the arrival of two consecutive photons, that is a photon arrival does not influence the dynamics of the next one. In a dissipative JJ (5) such reset occurs on a time scale $\sim 1/\alpha_J$, much shorter than the average arrival time $r_A^{-1}$ we are interested in. This is a decisive advantage for numerical simulations: for instance in the case discussed in Section~2.1, the value $r_A^{-1} \sim 1000\text{ s}$ is as high as $\sim 10^{12}$ normalized units, a number that can be challenging, to say the least, for nowadays stochastic numerical calculations. The quantity $r_0$ is essentially the Kramers rate, and therefore can be estimated analytically, for instance through Eq.~(14) and refined with Eq.~(12).

\begin{figure}[htbp]
\centering
\includegraphics[width=\columnwidth]{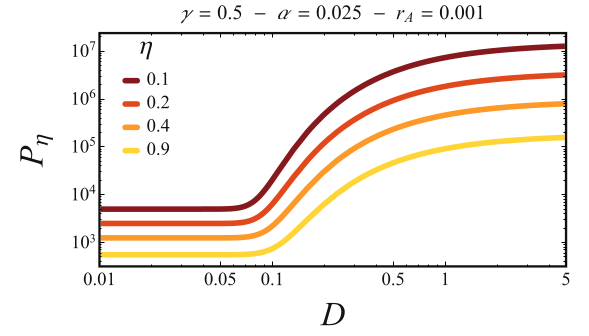}
\caption{Dependence of the observation time $P_\eta$, defined in Eq.~(34), on the noise amplitude $D$ and for different values of the quantum efficiency $\eta$. The other parameters are $\gamma_b = 0.5$ and $r_A = 10^{-3}$.}
\label{fig:fig5}
\end{figure}

Fig.~5 shows the dependence of the observation time $P_\eta$ on the noise amplitude $D$ for different values of the quantum efficiency $\eta$ and for fixed $\gamma_b = 0.5$ and $r_A = 10^{-3}$. We observe that $P_\eta$ shows a plateau for noise amplitudes below a certain threshold value, $D_{\text{th}}$. This limit noise amplitude can be calculated by imposing the condition $r_0(\gamma, D_{\text{th}}) = \eta r_A / 2$, from which one can obtain $D_{\text{th}} = \Delta U(\gamma) / \ln\left[\frac{\omega_a(r, \gamma)}{\pi \eta r_A}\right]$, where the attempt frequency is defined as $\omega_a = \sqrt{\frac{\alpha_J^2}{4} + \omega_0^2} - \frac{\alpha_J}{2}$. For $\gamma = 0.5$, $r_A = 10^{-3}$, and $\eta \in (0, 1]$ one obtains $0.06 \le D_{\text{th}} \le 0.12$.

Sample mean is by no means the best strategy for the detection. The best strategy is retrieved applying the maximum likelihood method \cite{addesso2012characterization}, which exploits the full information content of the escape distribution, and also gives the best results in a Bayesian analysis with equal probabilities assigned a priori. In particular, single photons can be analyzed with specially effective techniques that take advantage of the constant amplitude of the signals -- the amplitude $A$ in the photon train of Eq.~(8) is constant for quantum reasons. However, the analytical estimate of the distribution of the escapes is a relatively complicated problem \cite{lindner2004moments}; therefore a simple analysis has been preferred here.

\section{Conclusions}
In summary, a Josephson junction can be designed to detect a signal that mimics a train of single microwave photons, going beyond the condition of thermal noise much lower than the signal energy. An effective analysis methodology is described, showing that the detection is possible also when the Gaussian noise is comparable, and even slightly larger, than the photon energy. Such analysis is based on a heuristic SNR estimate through the Kumar-Carrol index, assuming a Poisson distribution of the events, and applying a sample mean statistics. Being the sample mean a suboptimal statistics, these results are to be interpreted as conservative. Better techniques could be devised in order to give a further boost to the quality of the detection. In details, the proposed analysis is based on the following assumptions:
\begin{itemize}
\item the Kumar-Carroll index is a good measure of the detector performances;
\item the sample mean is used instead of the full escape times distribution;
\item the Kramers rate, which is a good approximation only for relatively high barriers (respect to the noise energy), is used for all energies.
\end{itemize}
To verify these assumptions it is necessary to resort to numerical simulations or experimental verification of a JJ under the influence of a train of pulses that mimics the irradiation of photons.

Finally, a word of caution. In this work we do not give prescriptions for the junction fabrication neither for the cavity specifications, for we concentrate on the methodological aspects of photon detection through the analysis of the JJ switching times. Limitations in the fabrication process or in the electrodynamics of the cavity can restraint the detection performances. Further studies should be done to improve the model, in order to overcome the above limitations and to develop a full fledged detection strategy.

\section*{Declaration of Competing Interest}
The authors declare that they have no known competing financial interests or personal relationships that could have appeared to influence the work reported in this paper.

\section*{CRediT authorship contribution statement}
\textbf{A.S. Piedjou Komnang:} Conceptualization, Software, Data curation, Writing - review \& editing, Visualization. \textbf{C. Guarcello:} Conceptualization, Methodology, Software, Investigation, Data curation, Writing - review \& editing, Visualization. \textbf{C. Barone:} Conceptualization, Investigation, Writing - review \& editing. \textbf{C. Gatti:} Conceptualization, Writing - review \& editing. \textbf{S. Pagano:} Conceptualization, Investigation, Writing - review \& editing. \textbf{V. Pierro:} Conceptualization, Investigation, Writing - review \& editing. \textbf{A. Rettaroli:} Conceptualization, Writing - review \& editing. \textbf{G. Filatrella:} Conceptualization, Methodology, Formal analysis, Data curation, Writing - original draft, Visualization.

\begin{acknowledgments}
The authors wish to acknowledge financial support from Italian National Institute for Nuclear Physics INFN through the Project SIMP and from University of Salerno through projects FARB17PAGAN, FARB19PAGAN. GF and VP thank for support the Physics Department of the University of Salerno and INFN Gruppo Collegato Salerno.
\end{acknowledgments}

\end{document}